\documentclass[prb,aps,twocolumn,amsmath,amssymb,floatfix,showpacs,
superscriptaddress]{revtex4}

\usepackage[dvips]{graphics}
\usepackage{color}
\definecolor{dred}{rgb}{0,0,0.6}

\begin{document}

\title{\textcolor{dred}{Metal-insulator transition in an one-dimensional 
half-filled interacting mesoscopic ring with spinless fermions: Exact 
results}}

\author{Madhumita Saha}

\affiliation{Physics and Applied Mathematics Unit, Indian Statistical
Institute, 203 Barrackpore Trunk Road, Kolkata-700 108, India}

\author{Santanu K. Maiti}

\email{santanu.maiti@isical.ac.in}

\affiliation{Physics and Applied Mathematics Unit, Indian Statistical
Institute, 203 Barrackpore Trunk Road, Kolkata-700 108, India}

\begin{abstract}

We calculate persistent current of one-dimensional rings of fermions 
neglecting the spin degrees of freedom considering only nearest-neighbor
Coulomb interactions with different electron fillings in both ordered and 
disordered cases. We treat the interaction exactly and find eigenenergies 
by exact diagonalization of many-body Hamiltonian and compute persistent 
current by numerical derivative method. We also determine Drude weight 
to estimate the conducting nature of the system. From our numerical 
results, we obtain a metal-insulator transition in half-filled case with 
increasing correlation strength $U$ but away from half-filling no such 
transition is observed even for large $U$.  

\end{abstract}


\maketitle

\section{Introduction}

The metal-insulator (MI) transition is one of the most significant phenomena
in condensed matter physics~\cite{s1,s2,s3,s4,san1}. The usual band structure 
predictions are not capable of exploring many experimental evidences. 
For example, according to the band structure analysis, Iron(II) Oxide (FeO), 
Nickel Oxide (NiO) and Cobalt Oxide (CoO) are shown to be metallic but 
experimentally they exhibit insulating phase. Similarly, the absence of 
magnetism in several high-temperature superconductors verified in
experiments contradict theoretical results which suggest finite magnetism.
Likewise a strong mismatch has also been observed between theoretical
predictions and experimental observation in the determination of
band-gap for many semi-conductors~\cite{s4}. Considering Hubbard correlation, 
when people calculate the electronic band structures, it has flourished all 
the experimental results quiet strongly. If the Hubbard correlation is much 
greater than the bandwidth of a material, the MI transition becomes extremely 
relevant, and, compounds containing rare earth ions with localized $4f$ 
electrons or partially filled $d$-band elements show this type of transition. 
On the other hand, if bandwidth is higher than the Hubbard correlation, 
then the electron-electron interaction can be excluded and the experimental
evidences for those systems can be explained by conventional band structure
predictions~\cite{s4}. It is not only true for the macroscopic bulk systems 
but Hubbard correlation has an effective impact on low-dimensional transport 
phenomena~\cite{s5,s6}. 

Persistent current in an isolated $1$D small 
conducting ring threaded by a magnetic flux is a well established phenomenon 
in mesoscopic regime. There exist many controversial issues involving current
amplitude, sign and its periodicity between theoretical and experimental 
results~\cite{s5,s6,s9,s10,s11,s12,s13,s14,s15,s16,s17,s18,s19,s20,s21,s22},
and in last few decades people have attempted a lot to remove these 
discrepancies by improving their models~\cite{s5,s6,s20,s21,s22}. But still,
it is an open challenge to remove the disparity exactly. People have also 
explored the role of interactions on persistent current and to some extent
the enhancement of current has been observed in presence of 
interaction~\cite{s6}. In the early $90$'s Bouzerar {\em et al.} have 
performed the calculation of persistent current in $1$D ring considering 
spinless electrons. Utilizing a Wigner-Jordon transformation on the 
tight-binding (TB) Hamiltonian they have found the ground state energy 
from Lanczos algorithm and determined the current using conventional method.
It has been shown that the ring exhibits a MI transition at the half-filled 
band case due to correlation, and, away from half-filling the interaction 
has no such strong influence~\cite{s5}.
\begin{figure}[ht]
{\centering \resizebox*{4.5cm}{3.2cm}{\includegraphics{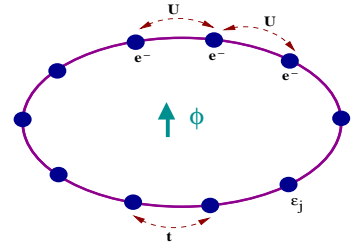}}\par}
\caption{(Color online). $1$D ring with interacting spinless electrons
threaded by an AB flux $\phi$ where the filled blue circles represent 
the atomic sites.}
\label{f1}
\end{figure}
Studying persistent current in finite two-dimensional ($2$D) 
arrays of semiconducting quantum dots in presence of magnetic field 
perpendicular to the $2$D plane the phenomenon of MI transition in the
half-filled band case has also been reported in a nice work~\cite{dassarma} 
by Das Sarma and his co-worker. However, the persistent current in $2$D 
quantum dot arrays has important differences compared to established persistent 
current in one-dimensional ring structure subjected to Aharonov-Bohm (AB) 
flux $\phi$. In that work (Ref.~\cite{dassarma}) several qualitative 
features of persistent current as functions of both on-site and extended
electron-electron interactions have been put forward where the current
has been computed by determining ground state energy through Lanczos
diagonalization technique. Here it is important to note that the
understanding of the MI transition in $1$D system of spinless fermions
in other context has already been established several years ago where 
response functions have been calculated~\cite{shankar} in a solvable model. 
These response functions, on the other hand, cannot be computed by Bethe 
ansatz though this ansatz allows us to get rigorous statements on 
many-body system of interacting $1$D spinless 
fermions~\cite{ansz1,ansz2,ansz3}.

The aim of this paper is to treat the interacting Hamiltonian {\em exactly} 
and find ground state energy from exact diagonalization of many-body TB
Hamiltonian. Evaluating ground state energy we compute persistent current 
using conventional method both for ordered and disordered cases. Our model 
comprises a $1$D finite ring of spinless fermions with nearest-neighbor (NN) 
Coulomb correlation. From our analysis we find that in the limit of 
half-filling current amplitude decreases sharply with correlation strength
$U$ and drops to almost zero (insulating phase) for moderate $U$, while a
metallic phase is always observed in the non-half-filled cases irrespective
of $U$. To substantiate the role of $U$ more precisely on conducting 
properties, in the present work, we also discuss the variation of Drude 
weight~\cite{s23} for different electron fillings, and, from these results 
sharp transition between two conducting phases is noticed. Finally, we 
discuss scaling behavior of persistent current with ring size.

The rest of the work is arranged as follows. Section II describes the 
model and theoretical approach and Sec. III presents the numerical 
results. Finally, we summarize our findings in Sec. IV.

\section{Model and Formalism}

The model is shown in Fig.~\ref{f1} where we consider an AB ring of
spinless fermions with nearest-neighbor Coulomb interaction. Under 
nearest-neighbor hopping approximation the TB Hamiltonian of such a 
$N$-site ring reads,
\begin{eqnarray}
\textbf{H} & = & \sum_j \epsilon_j n_j - t \sum_j \left[e^{i \theta} 
c_j^{\dagger} c_{j+1} + e^{-i \theta} c_{j+1}^{\dagger} c_j \right] +
\nonumber \\
& & + U \sum_{j=1} n_j n_{j+1}
\label{eq2}
\end{eqnarray}
where $c_j^{\dagger}$ and $c_j$ are the usual creation and annihilation
operators, respectively, $\epsilon_j$ describes the site energy, $t$ is 
the NN hopping integral and $U$ measures the NN Coulomb interaction. 
The phase factor $ \theta$ ($=2\pi\phi/N\phi_0$) arises due to AB flux 
$\phi$, where $\phi_0=c h/e$. For ordered ring, all site energies are 
identical and therefore we set them to zero without loss of generality. 
While, for disordered rings we choose site energies randomly from a `Box'
distribution function of width $W$ within the range $-W/2$ and $W/2$. 

To find ground state energy of the system first we construct the 
many-body Hamiltonian matrix considering the interaction exactly 
where the matrix elements are obtained following the prescription: 
$\textbf{H}_{mn} = \langle \psi_{m}|\textbf{H}|\psi_{n}\rangle$. Here
$|\psi_{m}\rangle$ and $|\psi_{n}\rangle$ are the basis vectors associated
with total number of spinless fermions $N_e$ in the ring. For example,
for a two-electron system we define them as
$|\psi_{m}\rangle=c_{p}^\dagger c_{q}^\dagger |0\rangle$ and
$|\psi_{n}\rangle=c_{k}^\dagger c_{l}^\dagger |0\rangle$ where $|0\rangle$ 
is the null state. Likewise we define basis vectors for higher-electron 
systems and construct appropriate matrices. Once the matrix is constructed,
we compute ground state energy by exact diagonalization method.

The persistent current in such a system at absolute zero temperature 
($T=0\;$K) can be determined from the relation~\cite{s18},
$I(\phi) = -\partial E_0(\phi)/\partial \phi$, where $E_0(\phi)$ is the 
ground state energy. 

Finally, we calculate Drude weight $D$ from the expression~\cite{s23},
\begin{equation}
D=\frac{N}{4 \pi^2} \left[\frac{\partial^2 E_0(\phi)}{\partial \phi^2}
\right]_{\phi \rightarrow 0}.
\label{equ4}
\end{equation}
Finite value of $D$ corresponds to the metallic phase while it drops to
zero for the insulating one, as originally put forward by Kohn.

\section{Results and Discussions}

Below we present our results. In our numerical calculations we choose
$c=e=h=1$ and measure the energy in unit of $t$ which is fixed at $1\;$eV.
Since we are dealing with exact many-body Hamiltonian, dimension of the
matrix increases sharply with electron filling, and therefore, we restrict
ourselves to the rings with few electrons due to computation limitations.
But, the point is that with these results we can easily analyze the
characteristic features of current as well as conducting properties for 
larger rings with higher $N_e$ as all the basic features will remain
invariant with our numerical results presented here.

\begin{figure}[ht]
{\centering \resizebox*{8cm}{6cm}{\includegraphics{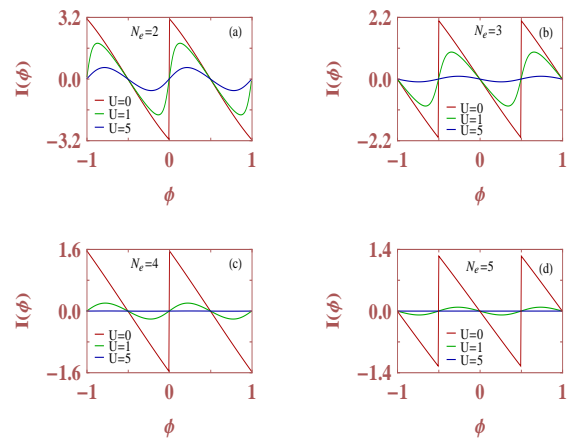}}\par}
\caption{(Color online). $I(\phi)$ vs $\phi$ characteristics for three
distinct values of $U$ in the ordered ($W=0$) half-filled ($N_e=N/2$) 
rings where (a), (b), (c) and (d) correspond to $N=4$, $6$, $8$ and $10$, 
respectively.}
\label{f2}
\end{figure}

Figure~\ref{f2} displays the current-flux characteristics of some typical
ordered half-filled mesoscopic rings for three distinct values of correlation
strength $U$ where (a), (b), (c) and (d) correspond to $N=4$, $6$, $8$ and
$10$, respectively. In the absence of Coulomb correlation, current exhibits
sharp transitions at $\phi=0$ (for even $N_e$) or $\pm 0.5$ (for odd $N_e$)
associated with the energy level crossing, while it becomes continuous as
long as interaction is included. Most interestingly we see that for a 
specific $U$ current decreases sharply as we increase ring size, and, for
\begin{figure}[ht]
{\centering \resizebox*{8cm}{6cm}{\includegraphics{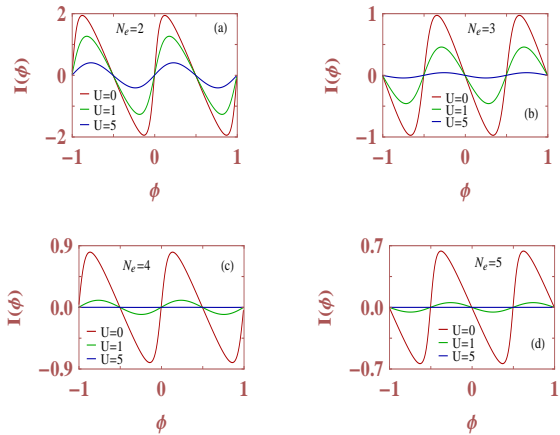}}\par}
\caption{(Color online). Current-flux characteristics in disordered ($W=2$)
half-filled rings for the identical parameter values taken in Fig.~\ref{f2}.}
\label{f3}
\end{figure}
large $U$ it practically drops to zero (blue line). This is solely due to 
the repulsive Coulomb interaction $U$. In the limit of half-filling all 
sites are occupied by single electrons and their hopping to the neighboring
sites strongly depend on the two physical parameters, viz, $t$ and $U$. 
\begin{figure}[ht]
{\centering \resizebox*{8cm}{6cm}{\includegraphics{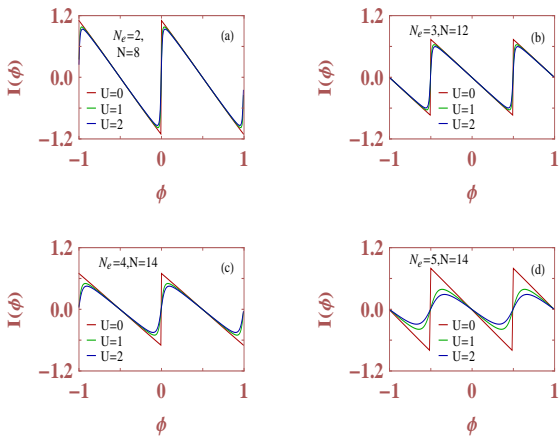}}\par}
\caption{(Color online). $I(\phi)$ vs $\phi$ characteristics for three
different values of $U$ in some ordered non-half-filled rings.}
\label{f4}
\end{figure}
For low $U$, when NN hopping dominates finite current is obtained, 
exhibiting a {\em conducting phase}. While for large $U$, the repulsive
interaction gets significant which suppresses electron hopping from one
site to the neighboring sites and it becomes more effective for larger 
ring. Therefore, current gets decreased with $U$ and with increasing ring
size it almost vanishes to zero which yields the {\em insulating phase}.

Certainly, further reduction of current is obtained when we include the
effect of disorder in such half-filled interacting rings as disorder
itself tries to localize the energy levels. The results of some typical
\begin{figure}[ht]
{\centering \resizebox*{8cm}{6cm}{\includegraphics{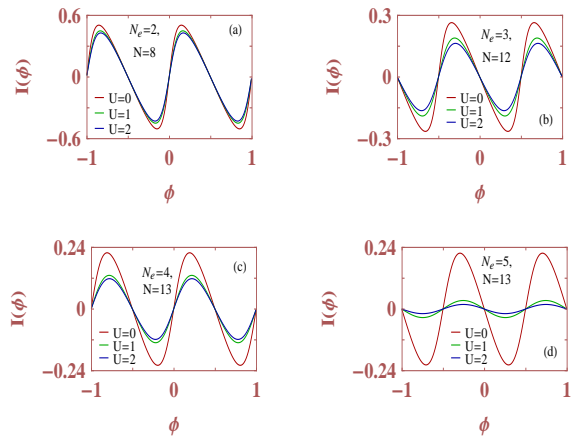}}\par}
\caption{(Color online). $I(\phi)$-$\phi$ characteristics for different
values of $U$ in some typical disordered ($W=2$) non-half-filled rings.}
\label{f5}
\end{figure}
disordered rings are shown in Fig.~\ref{f3}, where the averaging over 
fifty distinct disordered configurations are taken into account. Clearly
we see that a significant reduction of current takes place solely due to 
$W$, providing a continuous variation with $\phi$ because of the complete
\begin{figure}[ht]
{\centering \resizebox*{8cm}{6.5cm}{\includegraphics{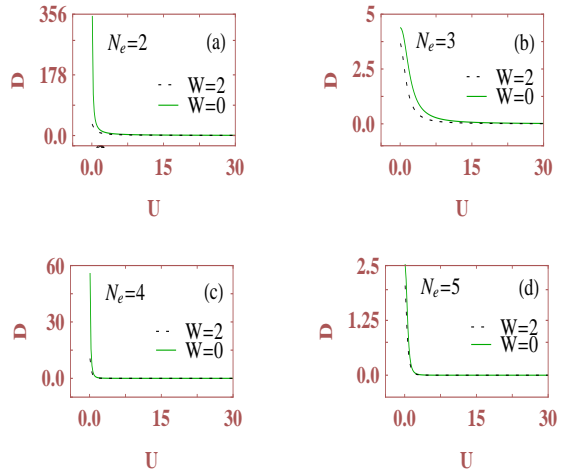}}\par}
\caption{(Color online). Variation of Drude weight $D$ (in unit of
$N/4\pi^2$) as a function of $U$ in the half-filled limit for both the
ordered and disordered rings.}
\label{f6}
\end{figure}
removal of degenerate energy levels, and it is further suppressed by the
repulsive factor $U$.

The situation becomes completely different when the rings are less than 
half-filled. To reveal this fact in Fig.~\ref{f4} we show the current-flux
characteristics of some typical ordered mesoscopic rings where we set
$N_e<N/2$. For these non-half-filled rings the rate of decrease of current 
with $U$ is too small and cannot be absolute zero even for a very large $U$.
The reason is that for such rings empty sites are always available where 
electrons can hop from the filled sites to these empty sites providing a 
\begin{figure}[ht]
{\centering \resizebox*{8cm}{6.5cm}{\includegraphics{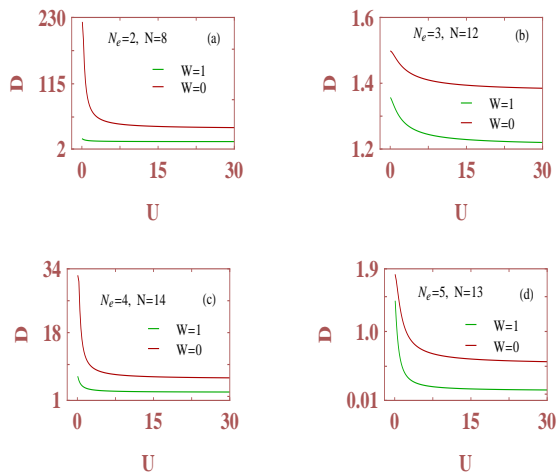}}\par}
\caption{(Color online). $D$ (in unit of $N/4\pi^2$) vs $U$ in the
non-half-filled rings for both ordered and disordered cases.}
\label{f7}
\end{figure}
\begin{figure}[ht]
{\centering \resizebox*{7cm}{6.75cm}{\includegraphics{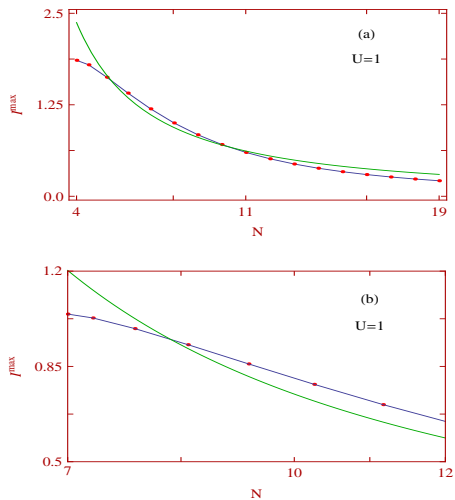}}\par}
\caption{(Color online). Variation of persistent current with ring size 
$N$ considering $U=1$ and $W=0$. The red dots, corresponding to the 
currents, are evaluated from our presented scheme and joining these
dots a continuous blue line is obtained. Using these dots we find a scaling
relation between the current and ring size which generates the green line.
Two cases are analyzed depending on $N_e$ where (a) and (b) correspond to
$N_e=2$ and $3$, respectively.}
\label{scl1}
\end{figure}
net circular current. Thus the rings exhibit only conducting phase and
insulating phase is no longer available when $N_e<N/2$, even for very large 
$U$.

Almost identical picture is also obtained even if we consider the effect of
disorder in these non-half-filled interacting rings. Though current gets 
reduced with $W$ itself, the existence of empty sites always provides a 
net current as clearly seen from the spectra given in Fig.~\ref{f5}.

The interplay between electron filling and Coulomb correlation on electronic
motion can be much more clearly explained by measuring Drude weight $D$, 
which essentially determines conducting nature of the system. In 
Fig.~\ref{f6} we show the variation of Drude weight $D$ as a function of $U$
for some typical half-filled rings for both the ordered and disordered cases.
Interestingly we see that for all these rings Drude weight falls sharply to
zero even for a slight increment of $U$ which reveals a crossover from the
conducting to the insulating phase of the system.

This insulating phase is no longer available even when the repulsive 
interaction is too high for less than half-filled rings. This behavior is
presented in Fig.~\ref{f7} where we plot the $D$-$U$ characteristics for
\begin{figure}[ht]
{\centering \resizebox*{7cm}{6.75cm}{\includegraphics{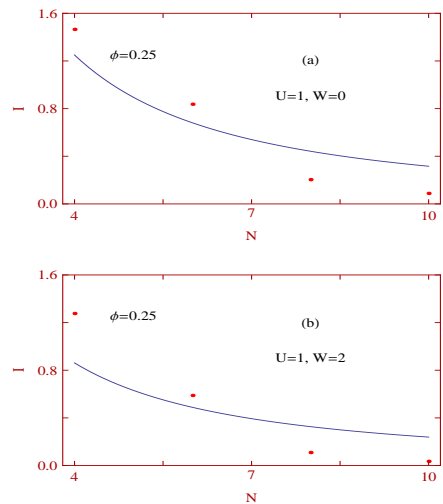}}\par}
\caption{(Color online). Dependence of persistent current with ring size 
$N$ in the half-filled band case when the Coulomb correlation strength is 
fixed at $U=1$. The red dots, corresponding to the currents, are evaluated 
from our presented scheme and using these dots we find a scaling
relation between $I$ and $N$ which produces the green line. Two cases are
analyzed depending on disorder strength $W$ where (a) and (b) correspond to
$W=0$ and $2$, respectively.}
\label{scl2}
\end{figure}
both the ordered and disordered rings, like above in Fig.~\ref{f6}, when
$N_e<N/2$. Though in presence of disorder electrical conductivity, viz, $D$
decreases but it never drops to zero, providing the metallic nature of the
system.

To make the present communication a self contained study now we focus our 
attention on the variation of persistent current with ring size $N$ and 
from it we try to figure out the scaling behavior.

In Fig.~\ref{scl1} we present the variation of $I^{\mbox{\tiny max}}$ 
with system size $N$ taking the maximum absolute value of persistent 
current from the current-flux curve (considering the variation of $\phi$ 
from $0$ to $\phi_0$). Two cases are analyzed depending on $N_e$, where 
(a) and (b) correspond to $N_e=2$ and $3$, respectively, and in both 
these cases we set $U=1$. The red dots,
representing the currents, are determined from our presented scheme
(described in Sec. II) and connecting these dots we get the continuous 
blue lines. With this dots we establish a scaling relation of the 
form: $I^{\mbox{\tiny max}}=CN^{-\xi}$ where the exponent $\xi$ becomes
$1.33$ both for $N_e=2$ and $3$ which we find from our extensive numerical
analysis. Whereas the pre-factor $C$ depends on $N_e$. It is $15$ for
$N_e=2$ and $16$ for $N_e=3$. This scaling relation matches very closely
to the previous analysis done by Gendiar {\em et al.}~\cite{gen} where they 
have shown by using bosonization techniques and additional approximations 
that persistent current decays algebraically with increasing system size 
and the exponent becomes exactly $1.33$ at half filling when $U=1$.

Finally, in Fig.~\ref{scl2} we show the dependence of
current with ring size $N$ in the half-filled band case setting the 
Coulomb interaction strength $U=1$. Since we restrict ourselves in the 
half-filled limit and due to our computation limitation, we cannot consider 
larger rings as the dimension of the matrix increases rapidly. Here, the 
red dots and the green lines represent the similar meaning as described 
in Fig.~\ref{scl1}. Two cases are analyzed depending on the disorder 
strength $W$, and, in both cases we find the identical scaling relation like
above i.e., $I \propto N^{-\xi}$, though the exponent changes with $W$.
For $W=0$ we get $\xi=1.5$, while it is $1.4$ for $W=2$. Most notably, from
our exhaustive numerical analysis we find that with increasing disorder
strength $\xi$ gradually decreases and eventually reach towards the 
limiting value $1.33$, which is consistent with the asymptotic behavior
as suggested by Gendiar {\em et al}~\cite{gen}.

\section{Summary}

To summarize, in the present work we have studied persistent current in a 
$1$D mesoscopic ring of spinless fermions with nearest-neighbor Coulomb 
interaction. Considering the interacting Hamiltonian {\em exactly} we have 
determined the energy eigenvalues through exact numerical diagonalization 
and computed persistent current using conventional derivative method. In 
addition we have also discussed conducting properties of the system by 
calculating Drude weight as a function of Coulomb correlation. From our 
analysis we have clearly shown that with increasing Coulomb correlation 
half-filled rings exhibit a metal-to-insulator transition, while, for 
less than half-filling only metallic phase is obtained irrespective of 
the strength of $U$. At the end, we have discussed scaling behavior of 
persistent current to address its asymptotic nature with ring size $N$.

\end{document}